\documentclass{article}

\usepackage{amsmath}
\usepackage{natbib}
\usepackage{todonotes}
\usepackage{comment}
\usepackage{fullpage}
\usepackage{paralist}

\usepackage{url}            

\DeclareMathOperator{\logonep}{log1p}
\newtheorem{definition}{Definition}
\newtheorem{example}{Example}
\newtheorem{test}{Test} 

\title{Guidelines for Implementing and Auditing Differentially Private Systems}
\author{
Daniel Kifer\thanks{Pennsylvania State University. \texttt{dkifer@cse.psu.edu}} \and Solomon Messing\thanks{Facebook. \texttt{solomon@fb.com}} \and  Aaron Roth\thanks{University of Pennsylvania. \texttt{aaroth@cis.upenn.edu}} \and  Abhradeep Thakurta\thanks{UC Santa Cruz. \texttt{aguhatha@ucsc.edu}} \and   Danfeng Zhang\thanks{Pennsylvania State University. \texttt{zhang@cse.psu.edu}}
}

\begin{document}

\maketitle
\begin{abstract}
Differential privacy is an information theoretic constraint on algorithms and code. It provides quantification of privacy leakage and formal privacy guarantees that are currently considered the gold standard in privacy protections. In this paper we provide an initial set of ``best practices'' for developing differentially private platforms, techniques for unit testing that are specific to differential privacy,   guidelines for checking if differential privacy is being applied correctly in an application, and recommendations for parameter settings. The genesis of this paper was an initiative by Facebook and Social Science One to provide social science researchers with programmatic access to a URL-shares dataset. In order to maximize the utility of the data for research while protecting privacy, researchers should access the data through an interactive platform that supports differential privacy.

The intention of this paper is to provide guidelines and recommendations that can generally be re-used in a wide variety of systems. For this reason, no specific platforms will be named, except for systems whose details and theory appear in academic papers.

\end{abstract}
\textbf{Funding Disclosure:} Authors Kifer, Roth, Thakurta, and Zhang were retained by Facebook as members of a ``Privacy Advisory Committee'' for the Facebook Election Research Commission (ERC) to help plan and evaluate some of Facebook's data releases, while Messing was a Facebook employee during initial but not final drafting. Facebook provided funding for this whitepaper and Messing contributed material to Section \ref{sec:usecase}. However, as an explicit condition of the engagement,  the Privacy Advisory Committee (not Facebook) retained full editorial control.

\section{Introduction}\label{sec:intro}
Datasets can provide a wealth of information to social scientists, statisticians and economists. However, data sharing is often restricted due to privacy concerns for users whose records are contained in the data. Differential privacy is the current gold standard in protecting their privacy while allowing for statistical computations on their data. It requires the careful addition of noise into computations over the data; the noise is designed to mask the effects of any one individual on the outcome of the computation.

The purpose of this whitepaper is to produce an initial set of ``best practices'' for testing, debugging, and code review of differential privacy implementations. To the best of our knowledge, such principles have not been codified before, but are necessary due to the increasing adoption of differential privacy for data sharing.

Data sharing organizations can provide differentially private access to pre-computed tabular data, as does the U.S. Census Bureau \citep{topdown}, or they can create \emph{systems} to allow external researchers to construct custom differentially private analyses. In this paper we focus on the latter, although the design principles can be applied to both approaches. 

Differential privacy platforms and languages \citep{pinq,Roy10Airavat,gupt,psidp,ectelo,Fuzz,Dfuzz,AFuzz,Fuzzi} simplify the construction of differentially private data analyses by providing API calls that allow users to specify the computations they want to perform from a limited menu. The platforms add appropriate amounts of noise into those computations and track the overall privacy risk of those data accesses. Granting the correctness of the platforms themselves, the correctness of the programs implemented within those platforms is guaranteed. 

Despite this obvious benefit, the real-world deployments of differential privacy to date have not made use of these systems, due to a number of limitations. First, the software engineering aspects of such platforms -- hardening against side-channels, support for unit testing, mature design principles, and code review standards -- are still lagging. In terms of side-channels,
many existing differential privacy systems have known vulnerabilities in their floating point implementations \citep{M12} and are susceptible to timing attacks \citep{Haeberlen:2011:DPU}. Furthermore, in complex systems it is easy to introduce subtle bugs that impact privacy leakage \citep{ebadistability}. Moreover because these systems seek to provide mathematical certainty about the correctness of the privacy properties of programs developed within them, they are necessarily limited to a set of predefined differential privacy primitives. Determining whether an arbitrary piece of code in a general-purpose programming language satisfies differential privacy is an undecidable problem, so systems that offer provable guarantees must pay a price in expressivity. 

There has been considerable progress in formally verifying the correctness of differential privacy programs \citep{Tchantz2011,Barthe12,BartheICALP2013,EignerCSF2013,Barthe14,EbadiPOPL2015,Barthe15,Barthe:2016:APC,Barthe16,langzero,Aws:synthesis,shadowdp}; however, many of these systems are designed to verify the correctness of differential privacy primitives, rather than to develop differentially private data analyses. Hence, they are difficult to use by data analysts and restrict the choice of programming languages and operations/libraries that they can use. A goal that is less ambitious than formal verification is debugging and testing. Research in testing differentially private programs \citep{Ding:2018:DVD,Bichsel:2018:DFD,gilbertproperty} is fairly new but these ideas inform much of our proposed test design. 

\subsection*{The role of testing.}
Verifying that a complex system satisfies differential privacy (or any other security property) is, in general, an intractable task,
and black-box statistical testing cannot in general positively establish privacy guarantees. One of the difficulties is that differential privacy requires a constraint to hold \emph{for every pair} of adjacent datasets (e.g., all possible pairs of datasets that differ on one record) and for every possible output of the system. Nevertheless, confidence in a system can be increased through a combination of 
\begin{inparaenum}[(1)]
\item  choosing a good set of design principles, 
\item performing code audits, 
\item developing a mathematical model of a system's semantics along with a proof of its correctness (possibly aided by software verification tools), and
\item designing (unit) tests that help ensure that the code conforms to the mathematical model. These tests can also help find subtle bugs in code/proofs and can find mismatches between implementation and pseudo-code.
\end{inparaenum}

The most actionable part of this paper is the testing framework it provides.
Thus, it is important to emphasize (again) that unit tests cannot guarantee that a system works correctly. At best, they can fail to prove that a system violates differential privacy.  In doing so, they can increase our confidence in the likelihood that a system designed to be differentially private in fact satisfies differential privacy. Unit testing is in no way a replacement for mathematical proof or formal verification, and the core belief that a system is differentially private should continue to stem from mathematical argument (either derived ``by hand'' or as a result of using a framework for differentially private programming that itself certifies correctness).

\subsection*{Organization}
The rest of this whitepaper is organized as follows. In Section \ref{sec:usecase}, we briefly describe the use-case that motivated the effort to establish these guidelines. In Section \ref{sec:dp} we present background information on differential privacy. In Section \ref{sec:system} we present recommended design principles for differential privacy platforms.  In Section \ref{sec:code} we discuss how code review can reduce the potential for privacy bugs and identify possible problems that code reviewers should check for.  In Section \ref{sec:test}, we describe methods for testing different components of a differential privacy system and in Section \ref{sec:summary} we present results and conclusions.

\section{Motivating Use Case}\label{sec:usecase}

In this section, we briefly describe the motivating use case and challenges encountered along the way. The project we describe is a work in progress, and the guidelines established in the remainder of this paper are intended to guide the development of an interactive system that best meets the needs of this use-case. It appears that in the short-term, this project will rely on a differentially private table release, while in the medium- and long-term it is possible that there will be an implementation of a differentially private query system to facilitate additional analyses and optimize the expenditure of privacy budget. 

\paragraph{Purpose and origin of the project.} In April 2018, an array of academic and funding organizations partnered with Facebook to make the company's data available to researchers to study the impact of social media on democracy and elections.\footnote{The partnership consisted of Social Science One (SS1), Facebook, the Social Science Research Council (SSRC), and the following foundations: the Laura and John Arnold Foundation, the Children's Investment Fund Foundation, the Democracy Fund, the William and Flora Hewlett Foundation, the John S. and James L. Knight Foundation, the Charles Koch Foundation, Omidyar Network’s Tech and Society Solutions Lab, and the Alfred P. Sloan Foundation.} The project was announced in the context of growing concern about both privacy and the use of the Facebook platform by nefarious actors to influence political outcomes. These concerns produced two countervailing pressures: protecting privacy on the one hand while providing data that yields as much precision as is possible.

After some deliberation, the parties decided to start by releasing a data set that allowed external researchers to study trends in misinformation from websites external to Facebook that were shared on its platform. Initially, the effort was to rely on K-anonymity to protect privacy. However, that approach was revised after the privacy community raised concerns that k-anonymity did not provide substantive privacy guarantees. The consortium started to explore sharing the data under differential privacy, which offers a substantially more rigorous guarantee (see Section \ref{sec:dp}). 

\paragraph{Data description.} The data that were initially collected were web page addresses (URLs) that have been ``shared'' on Facebook from January 1, 2017 up to and including August 6, 2019, along with actions users took (e.g., likes, clicks, etc.). URLs were included if the number of times they were shared (with public privacy settings) plus Laplace noise with scale 5 was larger than 100. URLs can encode private information, (e.g., http://social.security.numbers.com/query?bob=123-456-789), so Facebook sanitized them with various approaches such as removing common email address and phone number patterns \citep{fbURLSanitize}. As these URLs were already shared publicly by users, the URL strings were considered public knowledge (differential privacy was not applied to the set of web page addresses). The manipulations applied to URLs can therefore be seen as attempts to avoid amplifying mistakes that the sharers may have made (e.g., unintended public sharing).


The volume of the data makes a comprehensive ``static'' data release difficult (in particular, many end-users may not have the resources to process it). The data  include more than 50 TB per day for public sharing activity and interaction metrics, plus more than 1 PB per day for exposure data. The data were collected by logging actions taken on Facebook and processed using a combination of Hive, Presto, and Spark data systems, using the Dataswarm job-execution framework. In such situations, an interactive differential privacy system could be set up so that most of the computation associated with data analysis occurs on servers owned by an organization (e.g., Facebook or a third party) with sufficient processing capacity. A static data release, on the other hand, places the computational requirements of the data analysis on the end-user and provides less flexibility for the types of analyses that can be performed. However, it does ensure confidentiality of the analysis  (no one else knows what analysis the end-user is performing).

The underlying data consist of 3 tables: 
\begin{enumerate}
    \item A URL-table with descriptive information about URLs (actual URL, page title, brief description); the data in this table appears to be the least sensitive, since it it pertains to the URLs, and not the users' interactions with them (except, as discussed above, via the inclusion of the URL in the dataset).
    \item A user-URL-action table describing actions users have taken on specific URLs (likes, clicks, views, shares, comments, flagging for false news, etc.). This table contains sensitive data because whether or not a user has shared or liked a particular URL or group of URLs can reveal latent attributes, including sensitive personal characteristics (e.g. political affiliation, sexual orientation, etc.). 
    \item A user table describing the characteristics of the users in question. This data is also sensitive.
\end{enumerate}

\paragraph{Early problems.} Facebook sought to provide researchers with approved project proposal access to a differentially private query platform that would allow them to construct arbitrary queries (including joins) against these tables. At the time, however, pre-existing industrial-strength solutions did not meet the use case requirements. Some of these reasons include:

\begin{itemize}
\item Existing solutions were not able to cleanly provide privacy guarantees at the appropriate granularity. These solutions were tailored to tables that contain data pertaining to a single user in one and only one row. The data for this project contain user-url-action records, which are not useful if aggregated into a single record per user. 
\item The sheer size of our data meant that existing solutions were unable to scale to handle the necessary data operations.
\item Privacy guarantees are often difficult to assess in proprietary code, which complicates one of the desired features of differential privacy: mathematically-based privacy guarantees. This is especially true for less-mature codebases that do not implement clear testing standards. Open-source systems enable a larger community to vet the code. Furthermore, if code is open-sourced then end-users can, in principle, adjust their analyses for the privacy noise that is added to the computation. One of the benefits of differential privacy is that a correctly implemented system remains secure even when its source code is public.
\end{itemize}

\paragraph{DP table release.} 
As a stopgap measure, Facebook (and Social Science One) created a URL table and a table containing counts of actions taken related to each URL, broken down by user characteristics. 

Data aggregates that describe user actions were protected under a form of \emph{action}-level, zero-concentrated differential privacy (zCDP, see \citet{CDP}). The privacy parameter was chosen to be small enough to provide protections akin to user-level guarantees for 99\% of users in the dataset (i.e. 99\% of users in the dataset took sufficiently few actions that the action level guarantee provides a strong level of plausible deniability for all of their actions --- see Section \ref{sec:granularity} for further discussion of the important distinction between user and action level privacy).

User-level privacy guarantees are substantially more desirable than action-level guarantees, and action level privacy should be viewed as a principled compromise. Pragmatically, if user-level differential privacy  is not feasible, action-level guarantees may be a preferable alternative to complete suppression of the data (depending on its societal importance) and are similarly more preferable to data releases that use less rigorous privacy protections. An organization using action level privacy guarantees should not consider their job done, and should continue to improve their technology so that they can achieve user-level privacy for  subsequent data releases.


The release was a simple application of perturbation of the counts with Gaussian noise (these are often referred to as ``noisy measurements''). The release does not correct the data for consistency as, for example, the U.S. Census has done in its differentially private releases \citep{topdown}. That means (for example) that some values in DP-protected aggregated fields will be \textit{negative} due to the noise added, even though this is clearly impossible in the real data.\footnote{Note that truncating these counts at zero will bias statistical estimates.} However, this strategy allows researchers to compute unbiased estimates in a straightforward manner. In general, post-processing for criteria like consistency may destroy information (i.e., post-processing often cannot be inverted), and can always be performed by the end-user if desired. In almost all cases, a system should allow users to obtain such differentially private noisy measurements while giving them the option to obtain post-processed results as well. 

\paragraph{Going Forward.} In the future, we believe many organizations will be seeking to provide interactive access to their data via differential privacy platforms. 
Towards this end, the rest of this paper proposes design principles for such future systems.
This includes a particular focus on modularity and testing, so that the privacy guarantees of any future systems (which will inevitably be complex) can be rigorously evaluated and stress tested.

\section{Differential Privacy}\label{sec:dp}
In this section we define differential privacy and discuss what it does and does not promise, along with crucial decisions that a designer needs to make when deciding to use differential privacy. The suitability of differential privacy for an application depends on answers to questions such as: 
\begin{itemize}
\item Are the harms prevented by differential privacy of the same sort that we care about? (Sections \ref{sec:promised} and \ref{sec:notpromised}). 
\item At what ``granularity'' do we care about privacy? (Section \ref{sec:granularity}).
\item Does the privacy loss budget that we can ``afford'' in our application offer a reasonable quantitative guarantee? (Section \ref{sec:budget}). 
\end{itemize}
We note that for historical reasons we use the word ``privacy'' for what is often called ``confidentiality'' in the literature on law and ethics -- the protection of sensitive data/information. There, the term \emph{privacy} refers to an individual's control over their information and freedom from intrusion. 

\subsection{What Differential Privacy Promises}
\label{sec:promised}
Differential privacy requires algorithms that process data to be randomized -- given a certain input, we cannot predict with certainty what the output of an algorithm $M$ will be. However, for any possible input $D$ and output $\omega$, we may know the probability of the output $\omega$ occurring when $M$ is run with input $D$ -- that is, we may know the output distribution of $M(D)$. 

Differential privacy offers a very specific promise: if the data record corresponding to a single individual were removed from a dataset (or possibly changed in some other way --- see Section \ref{sec:granularity}), it would have only a controllably small effect on the distribution of outcomes for any analysis of the data. Formally, the mathematical guarantee is as follows. Let $X$ be an arbitrary data domain, and let a dataset $D$ be an unordered multiset of records from $X$.  We say that two datasets $D$ and $D'$ are user-level-\emph{neighbors} if $D$ can be derived from $D'$ by adding or removing the entire collection records associated with a single user. 
\begin{definition}[\cite{DKMMN06,DMNS06} User level Differential Privacy]\label{def:dp}
An algorithm $M$ mapping datasets to outcomes in some space $\mathcal{O}$ satisfies $(\epsilon,\delta)$ user-level differentially private if for all pairs of user-level-neighboring datasets $D, D'$ and for all events $E \subseteq \mathcal{O}$, we have that:
$$\Pr[M(D) \in S] \leq e^\epsilon\cdot \Pr[M(D') \in S] + \delta$$
\end{definition}
``User level'' differential privacy is the standard notion of differential privacy considered in the literature, and ``user-level'' is implicit when one refers just to ``differential privacy''. However, it is possible to define meaningful weakenings of user-level differential privacy such as ``action level'' differential privacy, which redefine the neighboring relation. See our discussion in Section \ref{sec:granularity}. We note at the outset that the privacy protections accompanying Facebook's release of the URL level dataset described in Section \ref{sec:usecase} are action level differential privacy, and \emph{not} user-level differential privacy.  

User level differential privacy has a number of interpretations. Here we list three, but see e.g. \cite{KS14,DR14} for a more in depth discussion. 
\begin{enumerate}
    \item Most directly, differential privacy promises that no individual will substantially increase their likelihood of coming to harm \emph{as a result of their data's inclusion in the dataset}. Said another way, for \emph{any event at all}, the probability that the event occurs increases or decreases in probability by only a little if their data is unilaterally added or removed from the dataset, and the remaining elements are left unchanged. Here, ``only a little'' is controlled by the parameters $\epsilon$ and $\delta$.
    \item Differential privacy controls the unilateral effect that any single person's data can have on the beliefs of any observer of the output of a differentially private computation, independent of what that observer's prior beliefs might have been. Specifically, for any outcome $o$, the beliefs of an outside observer after having observed $o$ given that an individual's data was included in the dataset are likely to be close to what they would have been had the outcome $o$ been observed, but the individual's data had not been included in the dataset. Here, ``close'' is controlled by $\epsilon$ and ``likely'' is controlled by $\delta$\footnote{This description most directly maps on to the guarantee of ``pointwise indistinguishability'' --- but as shown in \cite{KS14}, this is equivalent to $(\epsilon,\delta)$-differential privacy up to small constant factors.}.
     \item Differential privacy controls the ability of any outside observer to distinguish whether or not the unilateral decision was made to include your data in the dataset under study, holding the participation of all other individuals fixed. In particular, it implies that no statistical test aimed at determining, from the output of a differentially private computation, whether the input was $D$ or $D'$ (for any neighboring pair $D$ or $D'$) can succeed with probability substantially outperforming random guessing. Here ``substantially'' is what is controlled by the privacy parameters $\epsilon$ and $\delta$. It is this interpretation that is most actionable when it comes to testing: finding a statistical test for an algorithm $M$ that is able to reliably distinguish the distributions $M(D)$ and $M(D')$ for a particular pair of neighboring $D,D'$ (beyond what is consistent with the chosen parameters $\epsilon,\delta$) is enough to falsify a claim of differential privacy for the algorithm $M$. 
\end{enumerate}

These last two interpetations can be viewed as giving people a form of plausible deniability about the content of their data record. 

\subsection{What Differential Privacy Does Not Promise}
\label{sec:notpromised}

Differential privacy compares two hypothetical ``worlds'' $A$ and $B$. In world $A$, your data is used in carrying out some data analysis. In world $B$, your data is not used --- but the data analysis is still carried out, using the remainder of the dataset. If this data analysis satisfies differential privacy, then it guarantees in a strong sense that you should be almost indifferent between worlds $A$ and $B$ --- and hence if you do not fear any privacy harms in world $B$ (because your data was not used at all), you should also not fear privacy harms in world $A$. 

However, it is important to note that this guarantee --- although strong in many respects --- does not necessarily prevent many harms that might be colloquially thought of as related to privacy. Crucially, in both of the hypothetical worlds $A$ and $B$, the data analysis is carried out, and even with the guarantees of differential privacy, an individual might strongly prefer that the analysis had not been carried out. Here we discuss two stylized examples of inferences that are not prevented by differential privacy:
\begin{enumerate}
    \item \textbf{Statistical Inferences From Population Level Information}: The goal of statistical analyses is to learn generalizable facts about the population described by a dataset. By design, differential privacy allows for useful statistical analyses. However, inevitably, population level facts about the world, once learned, can be used in combination with observable features about an individual to make further inferences about them. For example, Facebook ``Likes'' turn out to exhibit significant correlation to individual features, like sexual orientation, political affiliation, religious views, and more \citep{KSG13}. These correlations can in principle be learned in a differentially private manner, and used to make inferences about these traits for any user who makes their "Likes" publicly observable. Note that the privacy loss to a hypothetical user in this situation here is not really due to the differentially private analyses that discovered correlations between Likes with sensitive attributes, but rather because the user allowed their Likes to be publicly visible. Nevertheless, this might feel like a privacy harm to a user who was not aware that the information that they were making public correlated to sensitive attributes that they did not want known.  
    \item \textbf{``Secrets'' embedded in the data of many people:} Differential privacy promises that essentially nothing can be learned from a data analysis that could not also have been learned without the use of your data. However, it does not necessarily hide ``secrets'' that are embedded in the data of many people, because these could have been learned without the data of any one of those people. Genetic data is a prime example of this phenomenon: an individual's genome, although highly ``personal'', contains information about close relatives. Similarly, on a social network, some kinds of information can be contained in the records of many individuals in the same social group: for example, a link or post that is widely shared. However, differential privacy \emph{does} protect information that is held amongst only a small number of individuals -- the ``group privacy'' property of $\epsilon$-differential privacy promises that the data of any set of $k$ individuals is given the protections of $k\epsilon$-differential privacy \citep{DR14}. 
\end{enumerate}

\subsection{The Granularity of Privacy}
\label{sec:granularity}
The discussion so far has been about the standard ``user-level'' guarantee of differential privacy: informally, as we discussed in Section \ref{sec:promised}, differential privacy promises a user plausible deniability about the entire content of their data records. However, it is also possible to define differential privacy at a finer granularity. Doing so offers a weaker but still meaningful guarantee. Formally, this is accomplished by modifying the definition of ``neighboring datasets'', which was defined (for the user-level guarantee) as datasets differing in the set of records associated with a single user. Other forms of finer granularity are possible, but here we mention two of particular interest:

\begin{itemize}
    \item \textbf{Action Level Privacy}: In some datasets, user records may simply be collections of \emph{actions} the users have taken. For example, in a Facebook dataset, actions might include pages and posts that the user has ``liked'', URLs she has shared, etc. Action level privacy corresponds to redefining the neighboring relation such that two neighboring datasets differ by the addition or subtraction of a single \emph{action} of one user. This correspondingly gives users a guarantee of plausible deniability at the level of a single action: although it might be possible to obtain confidence in the \emph{general trends} of a particular user's behavior, the user could continue to make statistically plausible claims about whether or not she had taken any \emph{particular} action.
    \item \textbf{Month (or Week, or Year) Level Privacy}: Similarly, in some datasets, user-level data might accumulate with time and be time stamped. This would be the case, for example, with actions in a Facebook dataset. Month level privacy corresponds to redefining the neighboring relation such that two neighboring datasets differ by the addition or subtraction of an (arbitrary) set of user actions all time-stamped with the same month. This correspondingly gives users a guarantee of plausible deniability at the level of a month: once again, although it might be possible to obtain confidence in long-term trends of the behavior of a particular user, the user could make a statistically plausible claim that their behavior during any particular month was whatever they assert. 
\end{itemize}

\textbf{Relationship to User Level Privacy:} The chosen granularity of privacy interacts with the quantitative level of privacy $\epsilon$ chosen via differential privacy's group privacy property. If e.g. a computation promises action level differential privacy at level $\epsilon = 0.001$, then it offers a guarantee similar to but still weaker than user-level privacy at level $\epsilon = 1$ for any user who has taken at most $1000$ actions. This guarantee degrades continuously with the number of actions taken: it offers a guarantee akin to user-level privacy at level $\epsilon = 2$ for any user who has taken at most 2000 actions, a guarantee akin to user-level privacy at level $\epsilon=3$ for users who have taken at most 3000 actions, etc. This guarantee is still weaker than user-level privacy because it does not prevent observers from inferring that power-users have 
taken very large numbers of actions (i.e. action-level privacy does not necessarily prevent inference about whether someone is a power-user or not). We note also that the quantitative guarantees of differential privacy degrade \emph{exponentially} with $\epsilon$ (due to the term $e^{\epsilon}$ in Definition \ref{def:dp}). Roughly, this means that under action-level differential privacy, a user who has taken $k$ times fewer actions is safer by a factor of $e^{\epsilon k}$. Although the scaling is approximately linear for values of $\epsilon < 1$, it starts degrading very quickly as $\epsilon$ becomes larger than $1$. In some settings, this guarantee might be acceptable when, for example, the vast majority of users in the dataset in question have taken fewer than $1000$ actions, and users that have taken many more actions are suspected of being accounts controlled by bots. Of course, in other situations, it might be precisely these users at the tails of the action distribution that are most in need of privacy, and which data analysts are most interested in --- in which case, action level privacy would not be a satisfying solution. In all cases, user level privacy is the stronger guarantee, but it will generally come at the cost of necessitating less accurate data releases.

\textbf{Potential Pitfalls of Action Level Privacy.} The choice of $\epsilon$ to use for action-level privacy often depends on how many actions most users take (e.g., 99\% of users take less than 1,000 actions). This statistic about actions taken cannot be obtained directly from the data (as this would constitute a violation of both user-level and action-level differential privacy). Instead, this number must be estimated using user-level differential privacy with techniques such as differentially private quantiles.

\textbf{Action-level Privacy and Data Manipulations.} Another potential risk of action-level privacy is that privacy guarantees can be inadvertently weakened by data manipulations such as adding records into a database. For example, if a dataset contains 1 million users, one can add 99 million ``fake'' users who have no actions at all. In such a case, all of the real users would move beyond the 99th percentile in terms of actions taken and hence would have degraded privacy guarantees. For this reason, organizations should strive to keep upgrading their privacy technology to user-level privacy protections.

\textbf{From Action Level to User Level Privacy via Truncation.} An alternative to action-level privacy is to pick a threshold (e.g., 1,000) and only keep at most 1,000 actions per user. This truncation threshold must be chosen without looking directly at the data. For example, it can be chosen a priori or  with the aid of user-level differential privacy algorithms (e.g., for quantile estimation). After the records are truncated,
user-level differential privacy can be applied to the resulting dataset, keeping in mind that neighboring datasets can differ by up to 1,000 records. The cost of this transition from action level to user level privacy is an increase in bias (since the activities of power users are truncated). If this bias is expected to be significant, one can use the threshold to divide the users into two disjoint sub-populations and provide differentially private statistics about each sub-population separately. A differential privacy platform can help with the privacy accounting.

\subsection{The Importance of the Privacy Budget}
\label{sec:budget}
A crucial feature of differential privacy is that it is compositional: If a sequence of $k$ analyses are carried out, such that each analysis $i$ is $(\epsilon_i,\delta_i)$-differentially private, then in aggregate, the entire sequence of analyses is $(\epsilon,\delta)$-differentially private for $\epsilon = \sum_{i=1}^k \epsilon_i, \delta = \sum_{i=1}^k\delta_i$ -- however, this is only a (often loose) upper bound. There are a number of other more sophisticated methods for ``accounting for'' the constituent privacy losses $(\epsilon_i,\delta_i)$ and aggregating them into an overall $(\epsilon,\delta)$ guarantee: but the important point is that 1) such aggregations of privacy loss are possible, and 2) the guarantees of differential privacy ultimately stem from the final $(\epsilon,\delta)$-guarantee characterizing the entire sequence of analyses performed. Differential privacy should therefore be thought of as a limited resource to parcel out,\footnote{This, in fact, is true of any privacy protection scheme that prevents reconstruction of the original database \citep{dinur:privacy}, not just differential privacy.} with some bound $(\epsilon_B,\delta_B)$ on the overall privacy usage. These parameters can be thought of as a privacy \emph{budget}. The choice of \emph{what} the privacy budget should be set at, and how it should be parcelled out are among the most important policy decisions that are made when deciding to allow for differentially private data access.

\paragraph{The Scale of the Privacy Budget}
In the following, we discuss the ``meaning'' of $\epsilon$ under the assumption that $\delta = 0$. In general, these interpretations will continue to hold with $\delta > 0$, with the caveat that the guarantees discussed may fail to hold with probability roughly $\delta$ (See \citet{KS14} for the formal derivation of this equivalence). We also note that mechanisms generally satisfy $(\epsilon, \delta)$-differential privacy for many different values of $\epsilon$ and $\delta$. Specifically, a mechanism $M$ generally has the property that for each value of $\delta$, there is a corresponding (possibly infinite) $\epsilon^\prime$ such that $M$ satisfies $(\epsilon^\prime,\delta)$-differential privacy. The function that relates $\epsilon^\prime$ and $\delta$ for the mechanism $M$ results in an $\epsilon, \delta$ curve. As long as this function always stays finite (i.e. for every $\delta$, the corresponding $\epsilon$ is not infinite), then $\delta$ should not be viewed as a ``catastrophic failure'' probability. Instead, a $(\epsilon^*,\delta^*)$ pair can be roughly interpreted as the statement that with probability at most $\delta^*$, the privacy guarantee that holds is weaker than $(\epsilon^*, 0)$-differential privacy.

The statistical testing interpretation of differential privacy implies the following. Suppose that an observer designs a statistical test for distinguishing whether an individual $i$'s data is contained in the dataset or not. If the test has significance level $\alpha$ (i.e., the risk of mistakenly concluding that the user is in the dataset when she is in fact not) then it cannot have power greater than $e^\epsilon\cdot \alpha$ (i.e., the probability of correctly concluding that the user is in the dataset when she is). So, for example, a test that has a false positive rate $\alpha$ of only 5\% will not be able to have a true positive rate that is greater than  $e^\epsilon\cdot 5\%$. For $\epsilon = 0.5$, this corresponds to a true positive rate just above 8\%. For $\epsilon = 1$, this corresponds to a true positive rate of 13\%. Note that the guarantee declines precipitously as $\epsilon$ increases beyond 1: When $\epsilon = 3$, this specific interpretation of the guarantee becomes almost vacuous. 

 For an alternative view, we can take the most basic interpretation of differential privacy: that events that have probability $p$ absent individual $i$'s data will have probability at most $e^\epsilon\cdot p$ when individual $i$'s data is included in the computation. Under this view, what a meaningful value of $\epsilon$ is depends on the estimate $p$ that one makes about the probability of the event that we are trying to prevent from occurring \emph{if the data analyses were run without individual $i$'s data}. Differential privacy guarantees that the event will occur with probability at most $e^\epsilon \cdot p$, and so a curator might be happy with values of $\epsilon$ such that $e^\epsilon \cdot p$ is deemed a manageable risk. We caution however, that it can be difficult or impossible to anticipate every possible harmful event that one might be worried about, or to estimate its probability $p$. 

More generally, one can attempt to choose the privacy budget by performing an economic analysis that trades off the privacy costs of the participants with the potential utility of an accurate analysis \citep{economic2,economic1,economic3}. This style of reasoning is delicate however and potentially brittle to modeling assumptions. 

\paragraph{Who Shares the Budget?}
In addition to deciding the numeric scale of the privacy budget $\epsilon_B$, there is another concern related to interactive differential privacy systems that provide API access to many users: how will the privacy budget be shared among a group of users. 
There are many choices that can be made and each one requires different tradeoffs. The tradeoffs to consider could include the strength of the final guarantee, and how much of that guarantee relies on the mathematics of differential privacy (as opposed to the contractual obligations of the users), together with the rate at which the ``privacy budget'' (and hence the useful lifetime of the dataset) will be expended.  

Two  of the most natural choices are the following:
\begin{enumerate}
    \item \textbf{One (Global) Privacy Budget}: The safest of the two (and the only one whose global guarantees rely only on the mathematics of differential privacy, rather than legal obligation) is to have a single privacy budget split among (and depleted via the actions of) all users. This offers the rigorous protections of differential privacy (and all of the interpretations discussed in Section \ref{sec:promised}) \emph{no matter how those users behave}, even if they aggregate all of their findings collectively, with the goal of trying to violate the privacy of a particular individual. In such cases, where the privacy budget is a shared resource, it often also makes sense to make the outputs of the system available to all users.    
    However, in applications in which many independent groups wish to have access to the same dataset for different purposes, we expect that this sharing approach will often exceed reasonable privacy budgets very quickly (since each group will need a minimal privacy budget level to ensure accuracy for some of their analyses). In this case it may be reasonable to consider an alternative solution.
    \item \textbf{Separate Privacy Budgets Across Organizational Structures --- with Contractual Protections}: One important use case is when there are many organizationally distinct groups of users (e.g. research groups at different universities, internal analytics teams in different parts of an organizational structure), etc. who may have no need or intention of communicating with one another over the course of their data analysis. If each of these groups is given their \emph{own} privacy budget, the guarantees of differential privacy (and all of the interpretations discussed in Section \ref{sec:promised}) will hold for each of these groups in isolation, \emph{so long as they do not share information with each other}. This guarantee relies on appropriately designed access control and may be acceptable, especially if additional legal protections are added on: for example, as a condition for obtaining access to the differentially private API, a research group may have to sign a contract forbidding them from sharing information obtained through the API with other research groups. In this case, the overall privacy guarantee will come from a combination of mathematical protections and legal protections. 
\end{enumerate}
The choice between the two options, of course, depends on an organization's policy and technical capabilities. An alternative to both of these options is a static data release. In such a case, the data curator should assemble a set of use cases and prepare a differentially private data release that satisfies most of them. 

\textbf{A Note on Cryptography:} Note that this problem is generally not solvable by the use of cryptography. For example, allowing computations to be performed on encrypted data, with the final result being decrypted by the user will generally not protect the data of individual users (unless the crypto protocols are paired with differential privacy). The reason is that cryptography does not provide any promises about what can be inferred about individual level data from information that is intentionally released. For protections against inferences, one still needs differential privacy. 




\section{System Design}\label{sec:system}
An interactive differentially private query answering system needs to be designed carefully to achieve the following (sometimes conflicting) goals:
\begin{itemize}
    \item \textbf{Expressivity}: The system should support a privacy-preserving version of common data analysis workflows, such as computing descriptive statistics (such as mean, median, variance of sub-populations), building machine learning models, and computing queries used for business decisions.
    \item \textbf{Modularity}: The system should be developed from sub-components that are easy to isolate from the rest of the system during testing. Differential privacy is a distributional property of a randomized computation. Thus, tests may need to execute some components millions of times in order to estimate the output distributions and verify that they meet expectations. This is made much easier if the individual components on which the claims of privacy rely are simple and fast to run. 
    \item \textbf{Minimality}: The privacy-critical components that need code review and testing should be few in number. This reduces the load on the code reviewer and provides fewer chances for information-leaking bugs to creep into the code base. It also increases the chance that the privacy-critical components can be formally verified by emerging program verification tools.
    \item \textbf{Security}: The system should prevent intentional and unintentional attempts at leaking more information than allowable by the differential privacy settings. This involves attention to details that stand outside of the abstract mathematical model in which differential privacy is usually proven: in particular, the details of random number generation, floating point arithmetic, and side-channels including system run-time.
    \item \textbf{Transparency}: A differentially private system provides approximate (noisy) answers to its users. The users need to adjust their statistical inferences based on the noise. Such adjustments are only possible if the users know exactly what was done to the original data. Moreover, a strength of differential privacy is that its  guarantees do not in any way rely on the secrecy of the algorithms used to analyze the data. As a result, ideally the code of a differentially private data analysis system will be open source and open to inspection. This also has the benefit of enabling a larger community to check for privacy errors, implementation bugs, thereby building trust in the system.
\end{itemize}

Modularity and minimality are possible because of two important properties of differential privacy: postprocessing immunity and composition. 

\begin{figure}[h!]
    \centering
    \includegraphics[scale=0.5]{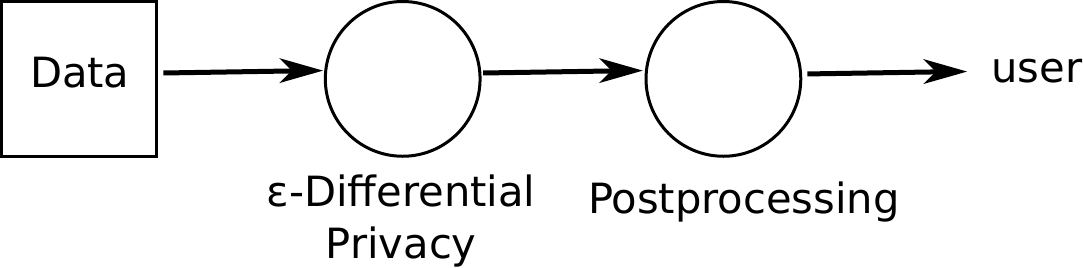}
    \caption{Postprocessing Immunity. If the first component satisfies $\epsilon$-differential privacy, then the entire workflow satisfies differential privacy with the same parameter $\epsilon$.}
    \label{fig:post}
\end{figure}

Postprocessing immunity, informally, means that we can do whatever we want with the output of an $\epsilon$-differentially private computation (so long as we don't make further use of the private data) without weakening its differential privacy guarantee. To illustrate this point, consider a simple algorithm that can be represented as a two-component flow diagram as in Figure \ref{fig:post}. The first component processes the sensitive data and satisfies $\epsilon$-differential privacy. Its output is fed into the second component. This is called the postprocessing component because it does not
directly access the data we want to protect -- it only processes its differentially private input. The entire algorithm then satisfies $\epsilon$-differential privacy with the same privacy loss budget (i.e., $\epsilon$) as the first component. In particular, this means that once we release differentially private data to the users, the users can do whatever they want with the data without compromising privacy (beyond the leakage allowed by the parameter settings). It also informs the system design that we recommend in Section \ref{sec:recommend} --- in particular, the ``postprocessing layer''. 

\begin{figure}[h!]
    \centering
    \begin{minipage}{0.45\textwidth}
    \includegraphics[scale=0.5]{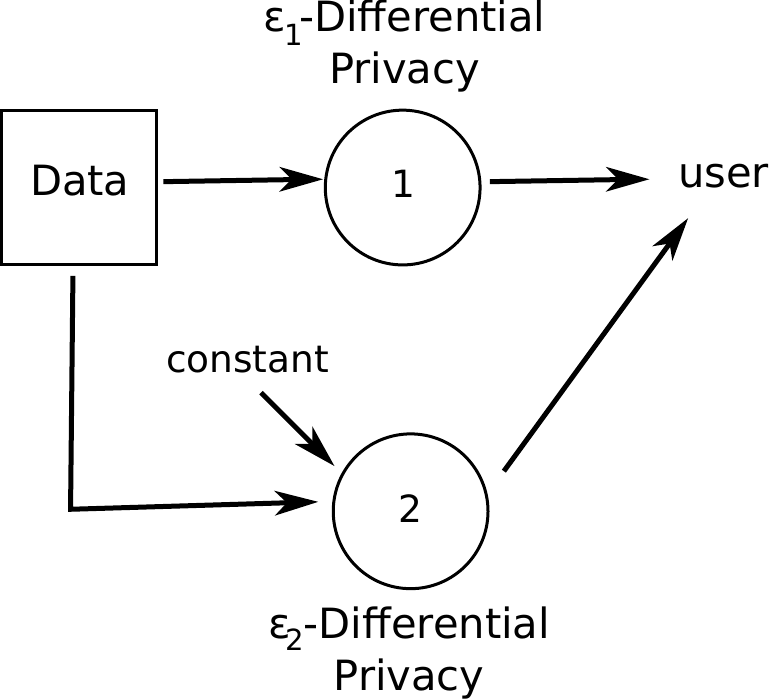}
    \caption{Independent composition. The total privacy guarantee is  $(\epsilon_1+\epsilon_2)$-differential privacy.}
    \label{fig:comp1}
   \end{minipage}$\qquad$
    \begin{minipage}{0.45\textwidth}
        \includegraphics[scale=0.5]{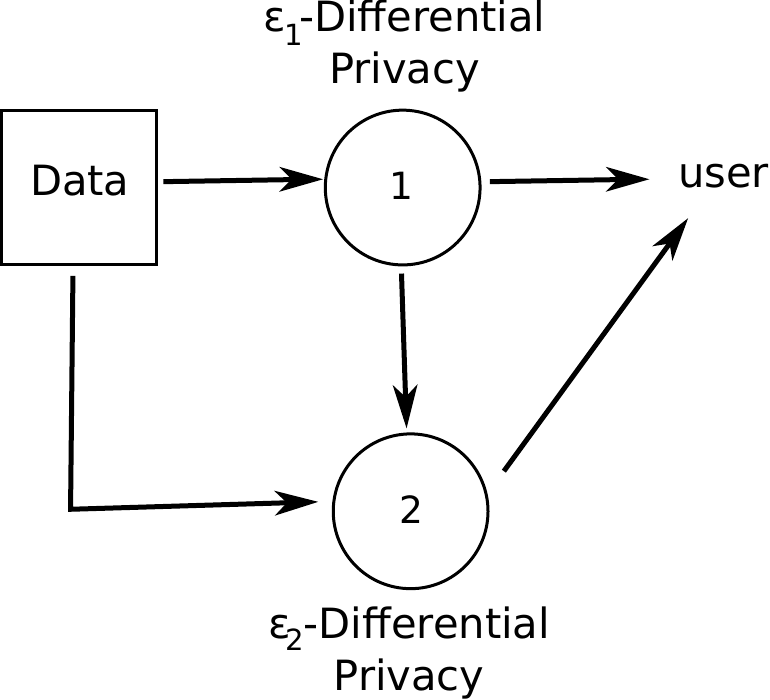}
    \caption{Sequential composition. The total privacy leakage is  $(\epsilon_1+\epsilon_2)$-differential privacy.}
    \label{fig:comp2}
    \end{minipage}
\end{figure}

On the other hand, composition related to the amount of information that is leaked when multiple differentially private processes are combined together. In the case of $\epsilon$-differential privacy (with $\delta=0$), the overall privacy loss is dictated by the sum of the individual privacy losses. Consider Figure \ref{fig:comp1}. Here the data are accessed by two independent differentially private algorithms. The first algorithm satisfies $\epsilon_1$-differential privacy while the second one (which possibly takes some constant as a second input) satisfies $\epsilon_2$-differential privacy. The combined release of both of their outputs satisfies $(\epsilon_1+\epsilon_2)$-differential privacy.

Now consider the situation in Figure \ref{fig:comp2}, where the constant input to Component 2 is replaced by the output of Component 1. This is a form of sequential composition. Releasing the outputs of both components to the user satisfies $(\epsilon_1+\epsilon_2)$-differential privacy. Note that generically, releasing only the output of Component 2 also results in $(\epsilon_1+\epsilon_2)$-differential privacy -- hiding the output of the first component does not necessarily reduce the privacy cost as it affects the operation of the second component.

There is a caveat in our discussion of sequential composition, which is that the privacy parameters $\epsilon_i$ are assumed to be fixed up front, before the computation begins. If Component 2 uses the output of Component 1 to dynamically determine its privacy level $\epsilon_2$, then this situation is known as \emph{parameter adaptive composition}. The simple composition theorem for pure differential privacy (i.e. $\delta=0$) can be  extended to this case \citep{RogersVRU16} without any loss in the composition guarantees, so that the overall privacy loss is the sum of the privacy loss of the two components. Similarly, ``additive'' composition theorems for approximate differential privacy (when $\delta > 0$) and variants like Renyi and concentrated differential privacy extend to this case. However, when $\delta > 0$, more sophisticated sub-additive composition theorems  do not necessarily extend to the parameter adaptive case, and adapting them can result in slightly larger overall privacy loss compared to the non parameter-adaptive case. See \citet{RogersVRU16,AFuzz} for more details. 

Finally, many differential privacy guarantees ultimately stem from a bound on the \emph{sensitivity} of a function. Function sensitivity measures the quantitative effect that the presence or absence of a single individual can have on the (exact) value of a function. Consider, for example, a simple function $f$, which, given a dataset $D$, computes $f(D)$ --- in our example, the number of users who have Facebook accounts and have listed their romantic preference as "Interested in Men". We would say that this function $f$ is ``1-sensitive'', because adding or removing a single user from the dataset $D$ could change this count by at most 1. More generally, the sensitivity of a 1-dimensional numeric valued function $f$ is defined as follows:

\begin{definition}
Given a function $f$ which takes as input a dataset $D$ and outputs a real number, its sensitivity is:
$$\Delta f = \max_{D \sim D'}|f(D) - f(D')|$$
where the maximum is taken over all pairs $D \sim D'$ of neighboring datasets.
\end{definition}
There are also multi-dimensional generalizations of sensitivity. Function sensitivity is important because it is often the primary thing we need to know about a function in order to use its value as input to a differentially private algorithm. For example, the simplest differentially private algorithms for estimating numeric valued functions (the ``Laplace'' and ``Gaussian'' mechanisms --- see \cite{DR14}) simply perturb the value $f(D)$ with noise with standard deviation scaled proportionally to $\Delta f$ --- the sensitivity of $f$. This also lends itself to a kind of modularity that informs our recommended system design in Section \ref{sec:recommend}: because function sensitivity can be tracked independently of differential privacy calculations as data is manipulated, and then passed to algorithms whose privacy guarantees depend on this sensitivity. 

\subsection{Recommended System Design}
\label{sec:recommend}

\begin{figure}[th!]
\centering
\includegraphics[scale=0.5]{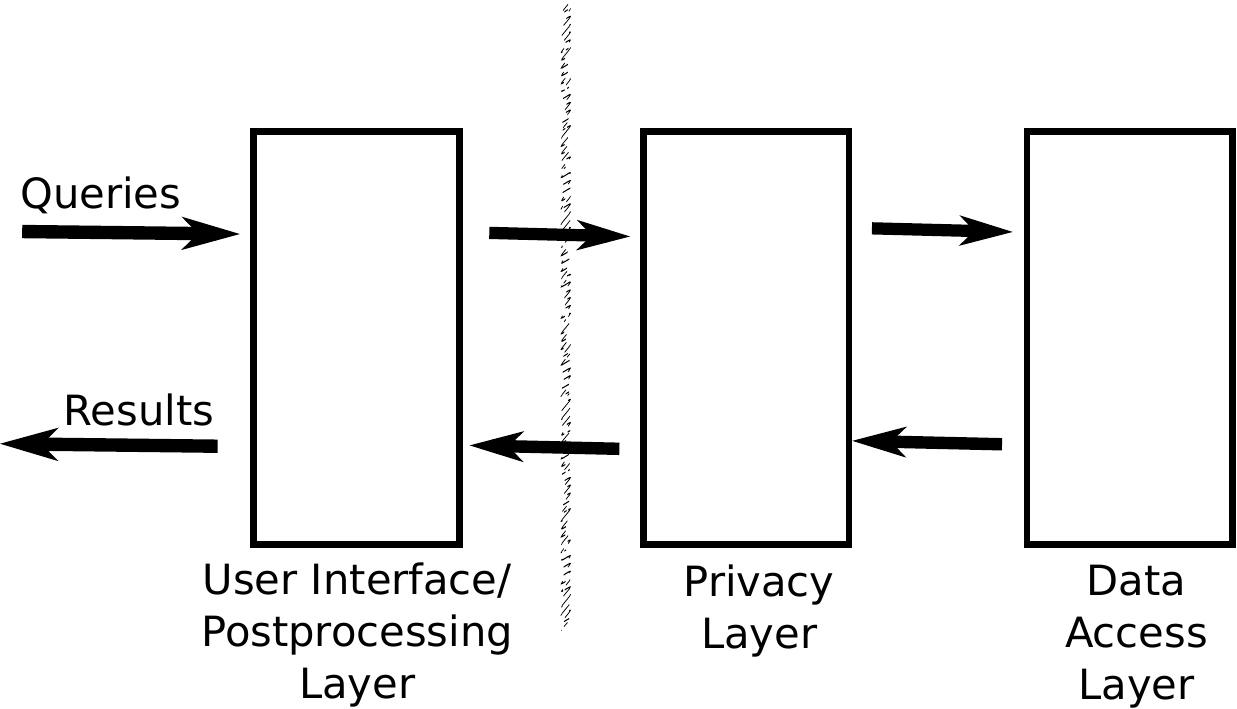}
\caption{Recommended System Architecture}\label{fig:arch}
\end{figure}

The principles of composition and postprocessing immunity together with the central importance of function sensitivity simplify the design of interactive differential privacy platforms to make it easier to achieve the desiderata listed at the beginning of Section \ref{sec:system}. A recommended system design, shown in Figure \ref{fig:arch}, consists of three layers: a data access layer, a privacy layer, and a postprocessing layer. Each layer performs a qualitatively distinct task, and the separation of a system into these three layers both lends itself to modular design, and simplifies the process of testing for correctness. 

\subsubsection{Data access layer.}
The data access layer is the only layer that should have access to the raw data. It performs transformations and aggregations on the data (e.g., joins, group by, and other SQL operations are simple examples). In general, it is this layer that performs exact computations on the data in response to user queries. For example, when computing a regression, functions in the data access layer might compute the covariance matrix of a set of datapoints, or when performing stochastic gradient descent to optimize model parameters in some class, the (exact) gradients of the current model would be computed in the data access layer. 

A crucial function of the data access layer is to compute (bounds on) the \emph{sensitivity} of the functions that it is computing on the data. Sensitivity calculations are deterministic, and can be automated under some circumstances: see for example the type system for tracking sensitivity developed in \citet{Fuzz} and subsequent work for the state-of-the-art in automating sensitivity calculations. 

Functions in the data access layer can be called only from the privacy layer, and they return to the privacy layer both an exact computation of some function on the raw data, and a bound on the sensitivity of that computation. 

\subsubsection{Privacy layer.}
The privacy layer is responsible for performing the randomized computations on functions of the data obtained from the data access layer that guarantee differential privacy at a particular level $\epsilon$. At a high level, the privacy layer has two distinct roles. First, it provides the implementations of the  ``base'' private computations that are supported --- e.g. simple counts and averages, estimates of marginal tables and correlations, etc. Second, it performs a book-keeping role: keeping track of a bound on the privacy loss that has accumulated through the use of these base computations. 

The base computations should take as input a desired base privacy parameter, and then make calls to the data access layer. Then, potentially as a function of the sensitivity bound provided by the data access layer, it performs the randomized computation that is guaranteed to protect privacy at the specified level. The base private computations are the cornerstone of the end-to-end privacy guarantee of the entire system. Hence, each of them should come with rigorous mathematical proof and/or formal verification to ensure that at algorithm-level, each base computation is deferentially private; moreover, testing and code auditing is required to validate the correctness of the implementation of each differentially-private algorithm.

Separately, the privacy layer keeps track of the total privacy cost expended during a sequence of analyses, and makes sure that it does not exceed some pre-specified privacy budget $\epsilon_B$. This is accomplished by keeping track of the privacy cost of each of the base computations, and applying an appropriate ``composition theorem'' to track the overall loss. In the simplest case in which the final guarantee is $(\epsilon_B,0)$-differential privacy, this is accomplished simply by summing the base privacy costs $\epsilon_i$ and making sure that $\sum_i \epsilon_i \leq \epsilon_B$ (and disallowing further computation if it will cause the privacy budget to be exceeded). However, more sophisticated methods can be employed when the desired final privacy guarantee is $(\epsilon_B,\delta_B)$-differential privacy for $\delta_B > 0$. This can involve tracking variants of differential privacy like ``Concentrated Differential Privacy'' \citep{CDP1,CDP} or ``Renyi Differential Privacy'' \citep{RDP}, which are measured and tracked in different units, but can be converted to standard $(\epsilon,\delta)$-differential privacy guarantees as desired (and can be used to ensure that the total privacy cost doesn't exceed a budget that is specified with standard parameters $\epsilon_B$ and $\delta_B$).

\subsubsection{Postprocessing layer.}\label{sec:design:post}
The postprocessing layer houses the user facing API, and makes calls to the privacy layer. Given queries supplied by the user, it verifies through the privacy layer that the desired computations will not exceed the privacy budget. If not, it makes queries to the privacy layer, and then potentially performs further transformations on them (the ``postprocessing'') before returning a response to the user. The guarantees of the privacy layer are sufficient to guarantee the privacy of answers returned through the postprocessing layer, because of differential privacy's postprocessing immunity property. 

As much computation as possible should be embedded in the post-processing layer so as to keep the set of base functions in the privacy layer as small as possible. For example, an algorithm for solving a linear regression problem privately might be implemented across layers as follows: the covariance matrix for a dataset would be computed in the data access layer, and then passed to an implementation of the Gaussian mechanism in the privacy layer (which would add noise in proportion to the automatically computed sensitivity of the covariance matrix). The privacy layer would return to the postprocessing layer a perturbed (and hence privacy preserving) covariance matrix. Only in the postprocessing layer would the covariance matrix be used to compute the final regression coefficients. 

The postprocessing layer should never have access to the original data. Hence, it should ideally be on a different server than the privacy and data-access layers. Aside from interfacing with the user, the purpose of the postprocessing layer is to keep the privacy layer as small as possible (to make it easier to verify and harder to attack). Because differential privacy is immune to postprocessing, under our suggested architecture, differential privacy guarantees will not be compromised even if there are errors in the implementation of functions in the postprocessing layer. Thus when implementing a new piece of functionality, designers should first carefully consider whether it can be implemented as a post-processing of functionality that already exists in the privacy layer (for example, if we already have code to compute a sum and a size, we do not need new code in the privacy layer to compute an average). This both keeps the attack surface as small as possible and reduces the burden of testing for correctness.  



\section{The Role of Code Review}\label{sec:code}

Differential privacy is ultimately a mathematical property of a system that must be proven in some idealized model --- and not something that can be established just through testing. The first step of designing a system should therefore involve specifying the idealized models in terms such as pseudo code and design document. The second step involves mathematical proof of the idealized models: that the sensitivity calculations from the data access layer are correct, that the base algorithms in the privacy layer have the privacy guarantees they claim to, and that the accounting for the privacy budget is correct. The role of code review is to check that the actual system as implemented conforms as closely as possible to the idealized models (in the form of pseudo code, design document etc.) in which privacy has been proven.

Our recommended system design simplifies this process by limiting the volume of code that has to be reviewed and compartmentalizing \emph{what} has to be reviewed:
\begin{enumerate}
    \item \textbf{Data Access Layer}: The only thing that needs to be reviewed at the data access layer are that the computed bounds on sensitivity correctly characterize the actual data transformations that are being done. There are existing type systems that can guarantee correctness for sensitivity calculations on a limited but expressive subset of data operations \citep{Fuzz}. Additional discussion about the documentation of the data access layer that needs to be checked can be found in Section \ref{subsec:access}
    \item \textbf{Privacy Layer}: Each base function in the privacy layer corresponds to an assertion that \emph{if} the inputs returned from the data access layer have sensitivity $\Delta$, then the privacy guarantee of the algorithm implemented in the base function is $f(\Delta)$, for some function $f$. It is this assertion that needs to be verified in the privacy layer --- which can be checked independently of the correctness of the sensitivity calculations in the data access layer. Typically, most base functions (such as Laplace mechanism, Exponential mechanism) come with rigorous mathematical proofs in the literature; in case of developing new base functions or variants of well-studied base functions, the developer should either provide rigorous mathematical proofs, or use verification tools to warrant their correctness.
    \item \textbf{Postprocessing Layer}: Functions in the post-processing layer have no bearing on the privacy guarantee offered by the algorithm. Hence from the perspective of auditing the privacy guarantee of a system, nothing in the postprocessing layer needs to be reviewed at all (other than checking that all data access to data goes through the privacy layer). This is the primary motivation for designing functionality so that as much code as possible is assigned to the postprocessing layer. 
\end{enumerate}

Code review for differentially private systems is not different in kind from code review for any other system aimed at guaranteeing that implemented code correctly matches a system specification --- except that it is more important, because differential privacy cannot be verified empirically. However, there are certain common pitfalls to watch out for, which we discuss in the following.

Naive implementations of differential privacy platforms are expected to have potential vulnerabilities in their source of randomness, their use of finite precision numbers, their use of optimizers, and in timing channels. Some of these vulnerabilities have published exploits and some do not. We note that implementations which defeat specific published attacks but do not solve the underlying problem are still considered vulnerable.

\subsection{The Source of Randomness}
The mathematical formulation of differential privacy requires perfect randomness \citep{DLMV12}. Perfect randomness is extremely difficult to achieve even in cryptographic applications, so real systems will have to settle for imperfect randomness. We can still ask for the same strong computational indistinguishability guarantees used for cryptography, however. Therefore we recommend that a code review check for the following: 

\begin{itemize}
    \item ``Random numbers'' should be generated by a cryptographically secure pseudo-random number generator (CSPRNG). Mersenee Twister is a commonly used pseudo-random number generator for statistical applications (in many libraries, it is the default) but is considered insecure for differential privacy.
    \item Seeds for the CSPRNG should not be hardcoded and should not be read from configuration files. Seeds should also never be saved.
    \item If two CSPRNG are used in parallel (for example in implementations using Hadoop, Spark, etc.) then they must explicitly be seeded differently. For example, one ``global'' CSPRNG can be used to seed the rest.
\end{itemize}

\subsection{Finite Precision Bugs}
Differentially private algorithms are typically analyzed in the idealized setting in which real-valued distributions (like Laplace and Gaussian distributions) can be sampled from, and in which all computations are carried out on real numbers (i.e. to infinite decimal precision).  In practice, however, these algorithms are necessarily implemented on finite-precision machines and this disconnect can lead to bugs that can leak significant amounts of information. Three important cases for a code review to focus on include random number generation, floating point computations that involve exponents and/or products, and continuous optimization solvers.

\subsubsection{Noise Generation.}

 \citet{M12} studied common implementations of the Laplace mechanism \citep{DMNS06,DR14} in a variety of languages and found that the Laplace distribution had ``holes'' corresponding to numbers that would never be generated. Such ``holes'' are shown to breach differential privacy since two adjacent datasets where random variables with these holes were added to finite precision numbers may miss different output values; hence the resulting mechanism cannot satisfy differential privacy. Paradoxically,  small privacy budgets  (e.g., $\epsilon=10^{-6}$), which in the idealized model would leak almost nothing were the ones that  resulted in the most leakage in practice. Similar issues can affect discrete distributions such as the discrete Laplace distribution \citep{Ghosh:2009:UUP} over the integers. This problem has not been fully solved but a variety of techniques can be used to mitigate the problems. These include:
 \begin{itemize}
     \item Use a secure version of a distribution that has been published in the literature. For example, the Snapping Mechanism \citep{M12} is a viable replacement for the Laplace mechanism.
    \item Place a lower limit on the privacy budget used for computations. Holes in distributions arise when the variance is large (and hence the privacy budget allocated to a computation is small). Placing a lower limit on the budget used in a mechanism can help mitigate this issue. For example, the Laplace mechanism adds noise with scale $\Delta/\epsilon$, where $\Delta$ is the sensitivity and $\epsilon$ is the budget allocated for this mechanism. As part of the mitigation procedure, one can require, for instance, that $\epsilon/\Delta\geq 10^{-3}$ to cap the variance and reduce the risk of holes.
    \item In addition to placing a lower limit on the privacy budget allocated to an operation, one can also discretize the output (for example, round the Laplace mechanism to the nearest integer).
 \end{itemize}
 
 Some mechanisms use continuous random variables to create discrete distributions. This includes selection algorithms such as ``Report Noisy Max'' \citep{DR14} and ``private selection from private candidates'' \citep{LT18}. Here the concerns of finite precision numbers are slightly different. Report Noisy Max takes a list of query answers as an input, adds Laplace noise to each, and returns the index of the query that had the largest noisy answer. The privacy analysis of such algorithms assumes that ties are impossible, which is true for continuous distributions but not for the samples produced by finite-precision computers. The true privacy guarantee is known as \emph{approximate differential privacy} \citep{DKMMN06}, which allows guarantees of differential privacy to fail with a very small probability. One concern in generating random variables inside these algorithms is that the chance of a tie should be minimized. In these cases, a standard floating point implementation of the Laplace distribution (with a lower bound on the privacy budget used) would be preferable to the Snapping Mechanism, as the Snapping Mechanism has a much higher chance of ties.

In other cases, discrete distributions are typically sampled in the following way (which is not recommended). Suppose the possible values are $a_1, \dots, a_m$ with probabilities $p_1,\dots, p_m$ (with $\sum_i p_i=1$). Form the cumulative sum: $c_j = \sum_{i=1}^j p_j$, draw a uniform random variable $U$ between $0$ and $1$, find the smallest $j$ such that $c_j\geq U$, and output $a_j$. This technique is commonly used for naive implementations of the exponential mechanism \citep{MT2007}, which samples from distributions of the form $\frac{\exp(b_i)}{\sum_j \exp(b_j)}$. Due to the exponentiation, some of the probabilities can underflow to 0, resulting in a potentially exploitable bug. It is recommended to avoid computation of probabilities as much as possible and to opt for mechanisms that use noise addition to achieve similar functionality -- for example, replacing the Exponential Mechanism with special versions of Report Noisy Max that use random variables drawn from the exponential distribution \citep{Barthe16}.

\subsubsection{Numerically Stable Computations.} 
 
Floating point operations such as addition differ from their mathematically idealized counterparts. For example, in pure mathematics, $A+B=A$ only occurs when $B=0$ and $(A+B)+C$ always equals $A+(B+C)$. These identities can fail in floating point. If $A$ is much larger than $B$, then $A+B=A$ even for nonzero $B$ due to rounding errors. There are three particular situations to look for in code review:
\begin{itemize}
    \item Multiplication of many numbers. The product $a_1\cdot a_2\cdot \cdots \cdot a_n$ is numerically unstable. If all of the $a_i$ are strictly between -1 and 1, this computation risks numerical underflow (resulting in 0). If all the $a_i$ have absolute value larger than 1, the product could result in an overflow (the program may treat it as $\infty$ or NaN). If some of the $a_i$ are close to 0 and others are large the result will depend on the order in which multiplication is performed (with possible answers ranging from 0 to $\infty$ depending on this order). Such multiplication often occurs when working with probabilities. It is better to work on a log scale, storing $\log a_i$ instead of $a_i$. Noting that $\log(a_1\cdot a_2\cdot \cdots\cdot a_n)=\sum_i \log(a_i)$, working in log scale results in significantly better numerical stability. Note that in this section, we take $\log$ to be the natural logarithm.
    \item Working with exponentials. Differential privacy often deals with quantities such as $e^{a}$. To avoid overflowing towards infinity, it is better to store $a$ instead of $e^{a}$ (again, this means working on a log scale).
    \item When working in log scale, one often needs to compute sums: we store $\log(a)$ and $\log(b)$ but need to compute $\log(a+b)$. The naive approach, which loses precision is to compute $\log(\exp(\log(a)) + \exp(\log(b)))$. A better approach is the following. Let $x=\log(a)$, $y=\log(b)$, $z=\max(x,y)$, and $v=\min(x,y)$. Then, mathematically $\log(a+b) = z + \log(\exp(x-z) + \exp(y-z)) = z + \log(1+\exp(v-z))$. We can further use the $\logonep$ function where $\logonep(c)$ is a much more accurate version of $\log(1+c)$ and is available in most numerical libraries (e.g., numpy). Hence we can compute $\log(a+b)$ as $z+\logonep(\exp(v-z))$ to obtain more precision than the naive approach for addition in log scale.
\end{itemize}

\subsubsection{Use of Continuous Optimizers.} 
Similarly, some algorithms have analyses in the idealized model that depend on computations being carried out to infinite decimal precision. An example is the  objective perturbation technique \citep{CMS11} often used for differentially private logistic regression. This technique  requires that an optimizer find the exact optimum of a convex optimization problem. In practice, such optimizations can only be approximated (since they are not guaranteed to have rational solutions in general) and often the optimization is stopped before reaching optimality (to reduce running time). Such inaccuracies must be accounted for in the privacy analysis. 

Some solutions to this problem include the following.
Objective perturbation and related algorithms could be replaced by a different method, such as certain types of output perturbation \citep{bolton}  that do not require optimality for privacy guarantees. Or, one could use a variant that does not require exact computation of the optimum \citep{approxOP1,approxOP2} -- these variants also perturb the output with a small amount of noise. 

\subsection{Use of Heuristic Optimizers}
There are a number of differentially private algorithms that make use of heuristic optimizers like integer program solvers \citep{dualquery,heuristics,approxOP2,topdown}. These heuristics may frequently work well, but because they are designed to try and solve NP-hard problems, they are not guaranteed to succeed on all inputs. This can make their use difficult to square with differential privacy (which is a worst-case guarantee). Thus, whenever heuristic solvers are used in differentially private systems, they should be a focus of the code review. Whenever possible, the privacy guarantees of the algorithm should not depend on the correctness of the solution found by the solver. To make this transparent, algorithms depending on solvers for utility guarantees but not for privacy should implement calls to the solver in the post-processing layer of the system \citep{dualquery}. This is by far the most preferred way to use a heuristic solver. 

When the correctness of the solver's output is important for the privacy guarantee of the algorithm \citep{approxOP2}, it is important that a solver be able to certify its own success (i.e. that it has returned an optimal solution to the optimization problem it was tasked with). Many commonly used mixed integer program solvers (e.g. Gurobi and CPLEX among others) can do this. Algorithms that use solvers with this guarantee can be converted into algorithms whose privacy guarantee does not depend on the success of the solver \citep{heuristics}, albeit with significant computational overhead. As an option of last resort, the reviewer should verify with extensive testing that the solver indeed successfully terminates reliably on a wide variety of realistic test cases.

\subsection{Timing Attacks}
Differential privacy is typically analyzed in an idealized model in which the only thing observable by the data analyst is the intended output of the algorithm. In practice, however, there are other side-channels though which information about the data may leak. Many of these side channels (e.g. electricity usage or electromagnetic radiation) are eliminated by making sure that user queries are executed on a remote server, but for systems designed to interactively field queries made by users, one significant one remains: execution time. The concern is that the following kind of adversarial query might be posed: ``given a boolean predicate $\phi$, return an approximate count of the number of data points $x$ such that $\phi(x)=\mathrm{TRUE}$''. This query has sensitivity 1, and so can be answered using standard perturbation techniques like the Laplace or Gaussian mechanisms. However, executing the query requires evaluating $\phi$ on every database element, and if evaluating $\phi$ takes much longer on certain database elements $x$ compared to others, then the running time of the query can reveal the presence of those elements in the database.

The best practice for closing timing channel is to make sure that --- as closely as possible --- queries to the dataset takes the same amount of time to complete, independently of the contents of the dataset. Guaranteeing this will typically involve both time-outs and padding: consider the example of estimating the number of database elements such that $\phi(x) = TRUE$, the system could evaluate $\phi(x)$ on each dataset element, and default to a value of ``TRUE'' if the execution of $\phi(x)$ took more than some pre-specified time $\xi$. Similarly, if $\phi(x)$ finishes evaluating in less time than $\xi$, then a delay should be added so that in total $\xi$ time elapses before moving on to the next element. $\xi$ should be chosen so that on almost all reasonable (i.e. non adversarial) queries, the timeout never occurs. If this can be accomplished, then the execution time of computing $\phi(x)$ does not leak any information about $x$. The running time of computing $\sum_{x \in D}\phi(x)$ will now scale linearly with $n$ (it will be roughly $n\cdot \xi$), which also leaks some private information through the total dataset size $n$. This can be mitigated by having the system expend a small amount of its privacy budget at startup to compute a private estimate $\hat n$ of the dataset size (which will also be useful for many other tasks, such as computing averages), and adding a delay to the overall run-time of the query that depends on $\hat n$, so that the run-time output itself is differentially private. For details of the implementation of timing attack mitigations, see \cite{Haeberlen:2011:DPU}. It is also possible to perform the paddings in an adaptive and automatic matter, by following the predictive mitigation mechanism~\citep{azm10,zam11}. Predictive mitigation is a general mechanism for mitigating timing channels in interactive systems: it starts with an initial prediction of computation time, and then for each query, it either pads the response time to the prediction (when $\phi(x)$ takes less time) or, updates the predication to be longer and pads the response time to the new prediction. Although predictive mitigation was designed for interactive systems in general, it can be adopted to mitigate timing channels in differential private systems.

Consultations with the NCC Group has raised additional indirect sources of timing attacks. For instance, if the function $\phi$ is allowed to use regular expressions, then an attacker can craft a regular expression that takes a long time to evaluate on a target record. Similarly, $\phi$ can create a large string (gigabytes in size) if a record is present. The resulting memory issues could slow execution and defeat improperly implemented timeouts.

\subsection{Running arbitrary code}
In principle, it is possible to allow users to create custom queries to be answered by writing an arbitrary piece of code. While this can be useful, it also poses special challenges that must be taken seriously. 

The first challenge is to bound the sensitivity of a user-provided function. Consider the following example: a hypothetical system allows the user to supply an arbitrary function $f$, written in a general programming language, that takes as input a database element $x$ and outputs a real number $f(x)$. The user would like to estimate the empirical average of this function on the dataset: $a = \frac{1}{|D|}\sum_{x \in D}f(x)$. In order to estimate this average privately (using say the Laplace or Gaussian mechanism), it is necessary to know the \emph{sensitivity} of $f$ which might be difficult to automatically bound if $f$ consists of arbitrary code (in general the problem is undecidable). One potential solution is to ask the user to specify that $f$ always returns values in $[\ell, u]$, and the system can instead estimate $\frac{1}{|D|}\sum_{x \in D}[f(x)]_{\ell,u}$, where the function $[r]_{\ell,u}$ clamps the value $r$ to always lie in the range $[\ell,u]$. The quantity $\sum_{x \in D}[f(x)]_{\ell,u}$ is guaranteed to have sensitivity bounded by $(u-\ell)$, and is equal to the quantity the user wanted to estimate if his assertion on the bounds of $f$ was correct. This is sufficient to guarantee differential privacy in the idealized model. 

The second challenge is that allowing the user to run arbitrary code opens up an attack surface for data to be exfiltrated through timing channels. For example, the function $f$ supplied by the user might attempt to set a value to a global variable as a function of the data point $x$ it is evaluated on (say, setting a global Boolean value to ``TRUE'' if a particular target individual's record is accessed, thus revealing through a side-channel that the target individual is in the dataset). If $f$ can be written in a general-purpose programming language, detecting such timing channels can be a very changing task, which has been extensively studied in the programming languages community~\citep{Agat00,Molnar:2005,zam12,pasareanu16,Chen:2017,Antonopoulos:2017,cached,ct-verif,caches,casym}. 

We recommend that either the system does not support writing user-defined queries of this sort, or else if it does, that the queries must be written in a small domain specific language such as Fuzz \citep{Haeberlen:2011:DPU} that is specifically designed to prevent side effects like this. See \citet{Haeberlen:2011:DPU} for further details of attacks that can result from running arbitrary code, and the mitigations that have been proposed.

\subsection{Peeking at the data}
It is important that the system does not peek at the data when making decisions (e.g., when choosing which functions to run or how much noise to use). All decisions must be noisy. Some common errors include revealing the exact attribute values that appear in a dataset (e.g., the exact set of diseases of patients in a hospital, the exact maximum age, etc.) and even the number of records in a table. Such deterministic information complicates privacy analyses \citep{constraints} and interferes with useful properties of differential privacy such as composition.

To see why revealing the number of records is problematic, consider a hospital with several different departments accessing the same underlying data. One wants to publish private statistics about cancer patients, another wants to publish statistics about patients with infectious diseases, and another wants to publish statistics about patients over 65. If each department publishes the size of view of the dataset, they are essentially publishing query answers with no noise. Pieces of information such as this can be combined together with additional differentially private statistics to further sharpen inference about individuals in the dataset \citep{KM2014}. Even if the publication of statistics only goes through one department in an organization, revealing the exact number of records should be avoided (because of future possible data releases by the organization or because of data releases by organizations about overlapping sets of people).

\section{Testing the System}\label{sec:test}
In addition to careful code review, it is important to set up an automated collection of unit-tests for privacy-critical components of the system. These tests cannot positively confirm that the system is differentially private, but can raise flags confirming that it is \emph{not}. Such tests are important even after a careful code-review because they can catch bugs that are introduced as the system is further developed. 

In a system that follows the recommended design, each layer has its own specific set of tests that can be run to help catch mistakes that could cause violations of privacy.

\subsection{Testing the Postprocessing Layer}

The postprocessing layer does not affect the privacy properties of the system if it is designed according to our recommendation, as it has no direct access to the original data. Thus standard penetration testing is sufficient. This is the reason why as much functionality as possible (like a function for ``average'' that just re-uses functionality in the privacy layer for ``sum'' and ``size'') should be implemented in this layer rather than in the privacy layer (as discussed in Section \ref{sec:design:post}) and why the postprocessing layer should be hosted on its own server.

\subsection{Testing the Privacy Layer}
The privacy layer consists of three main components:
\begin{itemize}
    \item The privacy accountant. The privacy accountant keeps track of the accumulated privacy loss of all queries made by the user, and makes sure that it does not exceed a pre-specified budget. For pure (i.e. $(\epsilon,\delta)$-privacy when $\delta = 0$) differential privacy, this amounts to keeping track of the sum of the privacy parameters $\epsilon_i$ spent on each operation and ensuring that the sum does not exceed the specified budget $\epsilon_B$. For approximate differential privacy $(\delta > 0)$ there are more sophisticated variants that may keep track of privacy in different units --- like  Renyi Differential Privacy \citep{renyidp} --- using  more sophisticated accounting procedures like the \emph{moments accountant} \citep{Abadi:2016:DLD}.
    \item Basic primitives that sample from distributions used by differentially private mechanisms (e.g. the Laplace distribution, the Gaussian distribution, the two-sided geometric distribution \citep{Ghosh:2009:UUP}, and the staircase distribution \citep{GKOV2015}). These functions do not access the data layer, but are implemented in the privacy layer because they are called by the differentially private mechanisms which do access the data layer, and their correctness is important for the final privacy guarantees of the system.
    \item Differentially private mechanisms (e.g., Report Noisy Max and Sparse Vector \citep{DR14}, the Laplace mechanism \citep{DMNS06}, and others) whose inputs are computed from the data and whose outputs are supposed to satisfy differential privacy. Note that the difference between the Laplace mechanism and the Laplace distribution is that the Laplace mechanism computes some quantity via a call to the data layer, and then perturbs it using a call to sample from the Laplace distribution (with parameters determined by the sensitivity of the data). The correctness of the Laplace mechanism depends both on the correctness of the function that samples from the Laplace distribution, and the correctness of the parameters it chooses as a function of the sensitivity of its input and the desired privacy parameter $\epsilon$. 
\end{itemize}

\subsubsection{Testing the privacy accountant.}
The privacy accountant accumulates the overall privacy impact of results returned to the user. To do this, it requires details about which mechanism was used and what data transformations were applied to the input data. The privacy accountant, however, cannot be given deterministic information about the original data (otherwise, such access would violate differential privacy). Validating the privacy accountant often requires three different steps. The first step is to make sure that  the privacy accountant is never accidentally bypassed -- it must record the impact of every privacy mechanism that was run. We note that this task can be aided by programming language support. For instance, by inheritance, we can enforce each mechanism to implement a privacy accountant method.

The second step is to make sure it can correctly measure the privacy impact of each mechanism in isolation. In the case of pure differential privacy, this is often very easy -- use the $\epsilon$ value reported by a mechanism. For more complex variants of differential privacy, such as Renyi Differential Privacy, the privacy impact is often provided in the literature as a formula that arises from evaluating an integral. In such cases, verifying the formula with numerical integration is useful. This is also a place where code review should be used to ensure the numerical stability of the floating point operations that are used.

The third step is to ensure that the total privacy impact is aggregated correctly. This involves taking a sequence of operations whose total privacy impact is already known (in the literature), and verifying that the privacy accountant claims the same (or larger) privacy leakage. In many cases, such as in the moment accountant, this simply involves addition of the privacy impacts of individual mechanisms (here it is also important to review the code for numerical stability). 

In some cases, testing the accumulated privacy impact is more complex. To illustrate this point, we consider how to account for linear queries. Suppose the input data can be represented as a vector $\vec{x}$ and suppose a change to one person's record can change this vector by at most 1 in the $L_1$ norm. That is, changing one person's record will result in a different vector $\vec{y}$ that is close to $\vec{x}$ in the following sense $\sum_i |y[i]-x[i]|\leq 1$. A linear query $\overrightarrow{q_i}$ can be represented as a vector having the same dimensionality as the data, so that $\overrightarrow{q_i}\cdot \vec{x}$ is the query answer.

\begin{example}
Suppose the vector $\vec{x}=(x[0], x[1], \dots, x[115])$ represents a histogram on age, with $x[j]$ being the number of people in the data whose age is equal to $j$.\footnote{Note that the upper bound on age is 115 in this example. This upper bound must be chosen without looking at the data. Any records with age larger than 115 must then be set to 115 (this process is called \emph{top-coding} or \emph{clipping}).} Adding or removing one person changes just one entry of the vector by at most 1 (so one person can change $\vec{x}$ by at most 1 in the $L_1$ norm). A query such as ``how many people are 18 and under?" can be represented as a vector $\overrightarrow{q_i}$, where $q[0]=q[1]=\cdots = q[18]=1$ while all other entries of $\overrightarrow{q_i}$ are 0.
\end{example}

Suppose we have $n$ queries $\overrightarrow{q_1}, \dots, \overrightarrow{q_n}$ and they are answered using Laplace noise with scale $\alpha_1,\dots, \alpha_n$ (i.e., for each $i$, we are given the noisy query answer $\overrightarrow{q_i}\cdot\vec{x} + \text{Laplace}(\alpha_i)$). To compute their precise privacy impact under $\epsilon$-differential privacy, we first define a matrix $Q$, where (for each $i$) row $i$ is the vector $\overrightarrow{q_i}$. We also define the diagonal matrix $D$, where entry $D[i,i]=1/\alpha_i$. For each column in the matrix product $D Q$ we can compute the $L_1$ norm of that column. Let $\ell$ be the largest $L_1$ norm among the columns. Then the release of the noisy query answers satisfies $\epsilon$-differential privacy with $\epsilon=\ell$ (but no smaller $\epsilon$ value).

Thus, when we check the privacy accountant (under pure differential privacy), it should produce an $\epsilon$ that is at least as large as the value $\ell$ computed above.

\subsubsection{Testing distributions.}

Testing the correctness of functions that sample from a distribution can be done using a goodness-of-fit test. For discrete distributions with a relatively small number of possible outcomes (e.g., 20), one can use the chi-squared goodness-of-fit test that is available in virtually any statistical software such as R, scipy, etc. Create a large number of outcomes (e.g., 10 million) and run the test.

For continuous one-dimensional distributions, a recommended goodness-of-fit test is the Anderson-Darling test \citep{andersondarling}. It is well-suited for differential privacy because it puts emphasis on the tails of a distribution (which is where privacy-related problems tend to show up). Suppose the distribution we want to sample from has a cumulative distribution function (CDF) $F$ and a density $f$. Let $x_1,\dots, x_n$ be points sampled from our code ($n$ should be large, like 10 million). The steps are as follows:
\begin{enumerate}
    \item First, sort the $x_i$. Let $y_1,y_2,\dots, y_n$ be those points in sorted order (so $y_1\leq y_2\leq \cdots\leq y_n$).
    \item Define the empirical distribution $G$ as follows: $G(t)=\frac{\text{number of $y_i$ that are $\leq t$\rule[-.3\baselineskip]{0pt}{\baselineskip}
}}{n}$
    \item Compute the test statistic $n\int_{-\infty}^\infty \frac{\Big(G(t)^2 - F(t)\Big)^2}{F(t)(1-F(t))}~f(t)dt$. For continuous distributions, this integral is equal to $-n - \sum_{i=1}^n \frac{2i-1}{n}\left(\ln(F(y_i)) + \ln(1-F(y_{n-i+1}))\right)$. 
    \item Compare the value of the test statistic to $3.8781250216053948842$, which is the 99\% percentile of the asymptotic distribution under the null hypothesis \citep{Marsaglia2004}. If the test statistic is larger than this number, the test has failed (it should only fail 1\% of the time). 
\end{enumerate}
In general, it is a good idea to use several different goodness-of-fit tests.

\subsubsection{White-box testing of privacy mechanisms.}
All differential privacy mechanisms used by a system should have an accompanying mathematical proof of correctness, or a proof from computer-aided verification tools with soundness guarantee. A reference to the literature is not always enough for the following reasons:
\begin{itemize}
    \item Some published mechanisms do not satisfy differential privacy for any $\epsilon$ due to errors that are discovered after publication.
    \item Some published mechanisms do satisfy differential privacy but have errors in the constants (i.e., they might under-estimate or over-estimate the privacy impact).
    \item Some published mechanisms are not accompanied by published proofs.
    \item Some published mechanisms may rely on assumptions stated elsewhere in a paper. Verifying their proofs is one way of identifying these assumptions.
    \item Some published mechanisms may use slightly different versions of differential privacy than being guaranteed by the system. For example, some variations consider the effect of \emph{modifying} one person's record (instead of adding/removing a record), or assume that a person can affect only one record in exactly one table of a multi-table dataset (which may not necessarily be the case with actual data).
\end{itemize}
Once a rigorous proof is established, one can design tests to determine that the code faithfully matches the pseudocode provided in the literature. For example, in a differentially private histogram  mechanisms, we may be able to analytically derive the variance of the histogram cells. This variance can be checked by running the mechanisms multiple times and empirically estimating the variance. 

This is another reason for the separation between the privacy layer and the data access layer. Creating an input histogram from raw data is usually an expensive operation. Turning the input histogram into a sanitized histogram that is a process that is much faster. Hence, for tests that require multiple runs of a mechanism, we want to avoid re-running slow components.

\subsubsection{Black-box testing of privacy mechanisms.}
In some cases, it may be difficult to derive properties of a privacy mechanism $M$ for white-box testing. In this case black-box testing can be done. The goal of a black-box test is to identify a neighboring pair of input databases $D_1$, $D_2$ and a set $S$ of possible outputs with the property that $P(M(D_1) \in S) > e^\epsilon P(M(D_2)\in S) + \delta$. If such $D_1, D_2,$ and $S$ are found, they constitute a counterexample that demonstrates violations of differential privacy. 

Searching for an appropriate $D_1, D_2,$ and $S$ is an area of active research and requires running the mechanism $M$ multiple times. Heuristic techniques for identifying promising choices for $D_1$ and $D_2$ are discussed by \cite{Ding:2018:DVD}. Typically, it suffices to use small input datasets (with few rows and columns). For example, to test one-dimensional mechanisms, like differentially private sums, variances, counts, and quantiles, one can first consider simple datasets having two columns (the first being a number ranging from 0 to 100 and the second from 0 to 1):
\begin{itemize}
    \item Dataset 1: $\emptyset$ (i.e., no records).
    \item Dataset 2: \{(0,0)\} 
    \item Dataset 3 \{(100, 1), (0,0)\}
    \item Dataset 4 \{(100, 1), (50, 0), (0,0)\}
\end{itemize}
Note that Datasets 1 and 2 are neighbors, so are 2 and 3, as well as 3 and 4. The feature of these datasets is that the records are nearly as different from one another as possible. In such cases, suitable choices for output $S$ could be an interval of the form $[a, b]$ or $(\infty, b]$ or $[a, \infty)$. In practice, one would evaluate many different neighboring pairs of databases and many different sets $S$ (e.g., many different intervals).

Once $D_1, D_2,$ and $S$ are chosen, verifying $P(M(D_1) \in S) > e^\epsilon P(M(D_2)\in S)+\delta$ can be done statistically by running the mechanism on both inputs many times and using a hypothesis test to determine if there is enough evidence to conclude that $P(M(D_1) \in S) > e^\epsilon P(M(D_2)\in S)+\delta$, as proposed by \cite{Ding:2018:DVD}. In some cases, the corresponding probabilities can be computed directly from the source code using tools such as the PSI solver \citep{psisolver,Bichsel:2018:DFD}. Many different values of $\epsilon$ should also be evaluated.

If counterexamples are found, the mechanism  does not satisfy differential privacy. If counterexamples are not found, then this adds more support to the working hypothesis that the mechanism is correct --- but does not constitute a proof of correctness. 

It is important to note that proper system design aids these black-box tests. Since they require running privacy mechanisms for millions of times, these tests benefit from a separation between the privacy layer and the data access layer (which houses slow data manipulation operations). Furthermore, the simpler the mechanism is, the easier to test. For example, suppose we have a two-stage mechanism for computing a differentially private mean. The first stage computes a noisy mean and the second stage forces the noisy mean to be nonnegative (i.e. a postprocessing stage). Black-box testing of the two stages together is less likely to successfully find counterexamples when only the first stage has bugs. However, if we move the postprocessing stage to the postprocessing layer (where it belongs), then we just need to test the first stage by itself, making it easier to discover if it has any errors.

\subsection{Testing the Data Access Layer}\label{subsec:access}
The data access layer is responsible for managing and querying the original data. Hadoop, Spark, Hive and relational database are backends that can be part of this layer. While it is performing data transformations (such as selecting rows, dropping columns, aggregating values), it needs to keep track of quantities such as:
\begin{itemize}
    \item The set of valid values for a transformation. For example, if we group records by disease, we need to keep track of what the valid groups (diseases) are. These have to be pre-defined (not taken from the data). For numeric attributes, this could include computation on the bounds of attributes. For example, there need to be pre-defined upper and lower bounds on Age. If we are interested in squaring the ages, the system needs to keep track of upper and lower bounds on the square of ages (these bounds cannot depend on the data).
    \item Sensitivity and stability. One can often define a distance over datasets  that result from data transformations. For transformations that return datasets (e.g., the SQL query ``SELECT Age, Income FROM Salaries WHERE Income > 30,000"), one natural metric between two datasets $A$ and $B$ is the size of their symmetric difference $|A \ominus B|$ (how many records must be added to or removed from $A$ to obtain $B$). For transformations that return vectors of numbers, a natural metric is the $L_1$ norm (common for pure differential privacy) or the $L_2$ norm (common for Renyi Differential Privacy \citep{renyidp}). These metrics help us quantify the possible effect that one record could have on the output of a transformation. For numerical transformations, the sensitivity measures (using $L_1$ or $L_2$ distance)  how much the output of a transformation can be affected by adding or removing one record in the worst case. For transformations that output datasets, the corresponding concept is called \emph{stability} \citep{pinq,ebadistability} and uses the size of the symmetric difference to quantify how much the output can change.
\end{itemize}
Thus testing the data access layer requires that it produces the correct query answers, that it properly keeps track of the data domain (and/or bounds on the data), that it properly computes sensitivity, and that it properly computes stability. Note that domain, bounds, sensitivity, and stability can only depend on public knowledge about the data (such as the database schema) but cannot depend on any private information (e.g., the actual records).

The intermediate data products produced within this layer generally have at least three types such as:
\begin{itemize}
    \item Datasets -- tables with rows corresponding to records and columns corresponding to record attributes. An example is an employees table that records the name, title, workplace location, and salary of every employee in a company.
    \item Grouped datasets. Suppose we group the records of this employees table by title and workplace location. Every record in the resulting table consists of a title (e.g., "manager"), workplace location (e.g., Menlo Park), and a set of employee records who match the title and workplace location (i.e., the records of managers in Menlo Park). Thus this grouped table has the following columns: ``title'', ``workplace location'', and ``record-set''. 
    \item Vector -- a list of predefined data statistics. For example, a two-dimensional vector could include the most frequent title as the first dimension and the overall average salary in the second dimension. We can also think of a scalar as a vector of dimension 1.
\end{itemize}
The reason for grouping data products into types is that different operations are available on different data types. For example we can perform SQL queries on tables, we can additionally perform aggregations on each record in a grouped dataset, and we can perform vector operations on a vector. Furthermore, for each type we may need to store different information that summarizes the worst-case influence that adding or removing one record can have.

\subsubsection{Tracking Bounds and the Data Domain}
For each column in a table, we must keep track of the valid values for that column. These values cannot depend on the data. For instance, if a hospital produced a  medical records table for patients with infectious diseases, the set of diseases must be predefined (for instance, ``ebola'' should be one of those diseases even if no patients at the hospital ever contracted ebola). For numerical attributes, one can maintain upper and lower bounds on the attributes (e.g., a lower bound on recorded age and an upper bound on recorded age). All of this can be considered metadata.

For data access operations that output tables, unit tests must ensure that the metadata does not depend on the actual data records. For example if a system top-codes age at 115 (i.e. ages over 115 are converted to 115), then the age column must have an upper bound of 115 even if such ages do not appear in the data. Similarly, suppose ``Workers'' is a table that keeps track of the name, title, and age of each worker. Suppose the \emph{a priori} bounds on age in this table ensure that it is between 18 and 65 (inclusive). If we perform the SQL query such ``SELECT * FROM Workers WHERE Age $< 25$'' then upper and lower bounds on the resulting table should be 25 and 18, respectively, even if this table is empty.

As tracking bounds and allowable sets of values can be complex, it is advisable that system designers create a document detailing how each operation affects the allowable domain and attribute bounds for its input tables. This can be augmented by runtime assertions. For example, if the system expects that age is between 18 and 25 in a table that it has produced, it can go through each record to ensure that the bounds hold. What should the system do if it finds a violation of the bounds? It must not throw an error or perform any special action that is detectable by a user (as this would leak information). Instead, it should correct the violating record (e.g., trim the data) so that it satisfies the bound and log an error message that can only be viewed by system developers (since a bug has been detected). The logging mechanism must not introduce side channel vulnerabilities (e.g., it should not delay system processing by a noticeable amount, or noticeable increase memory usage).

It is important to note that data transformation operations such as GroupBy can differ from traditional implementations in databases and big-data platforms such as Spark. If we have a table of infectious diseases and we perform a GroupBy operation on disease, the resulting grouped table should have a entry for each possible disease (not just diseases that appear in the data). It is important to specifically generate unit tests for this case.

\subsubsection{Tracking Stability}\label{subsec:stability}
For operations whose inputs are tables (or grouped tables) and outputs are also tables (or grouped tables), the system  must track how a change to the input propagates to the output. For example, suppose a user asks the system to perform the following sequence of operations on an Employee table:
\begin{itemize}
\item Temp1 = (SELECT * FROM Employee) UNION (SELECT * FROM Employee)
\item Temp2 = (SELECT * FROM Temp1) UNION (SELECT * FROM Temp1)
\item Temp3 = (SELECT * FROM Temp2) UNION (SELECT * FROM Temp2)
\item Temp4 = (SELECT * FROM Temp3) UNION (SELECT * FROM Temp3)
\item Temp5 = (SELECT * FROM Temp4) UNION (SELECT * FROM Temp4)
\item SELECT SUM(Salary) FROM Temp5 WHERE Employee\_id==`1234567'
\end{itemize}
Suppose salaries are capped at \$300,000. How much noise should be added to the result? One could reason holistically about the whole set of queries and determine that the largest change in the answer occurs from adding one person with salary \$300,000 to the input table (causing 32 records to be added to Temp5). Hence, if using the Laplace mechanism, Laplace noise with scale $(300,000 \cdot 32/\epsilon)$ is needed. In general, it is difficult to program this kind of holistic reasoning into a system but it is easier to track the effect of each transformation in isolation and then to combine their results. For datasets and grouped datasets, this is tracked using the notion of \emph{stability} of an operation $f$, which depends on the concept of symmetric difference. Given two tables $A$ and $B$, their symmetric difference $A \ominus B$ is the set of records that appear in one, but not both of the datasets (i.e. the records that appear in $A\cup B$ but not in $A\cap B$). The stability of a function is a rule that explains how a perturbation in the inputs (as measured by the size of the symmetric difference) affects the output. For instance, let $f$ be the union operation on two datasets $A$ and $B$ that may contain the same people. If we have two other datasets $A^\prime$ and $B^\prime$ such that $|A\ominus A^\prime| = a$ and $|B\ominus B^\prime|=b$ then clearly $|f(A,B)\ominus f(A^\prime, B^\prime)|\leq a+b$. In other words, if we add/remove a total of $a$ records to $A$ and a total of $b$ records to $B$, we can change the union by at most $a+b$ records. By applying this rule sequentially to the above sequence of operations, we can track the influence of one person as we are performing the computation: adding/removing 1 person to Employees affects Temp1 by at most 1+1=2 records, which affects Temp2 by at most 2+2=4 records, etc.

Thus unit tests for stability of an operation $f$ involve creating datasets, measuring the size of their symmetric differences, and then measuring the sizes of the symmetric differences after applying an operation $f$ to each. For some common cases, the stability is given in the literature \citep{ebadistability}:
\begin{itemize}
    \item GroupBy has a stability of 2. Adding or removing a total of $k$ records in the input causes at most $k$ groups to change. This results in a symmetric difference of size $2k$ (since changing one group is the same as removing that group and then adding the new value for that group).
    \item Many query languages have a ``limit m'' operation which returns at most $m$ records from a dataset. Adding or removing a total of $k$ records can cause the size of the symmetric difference to change by either $2k$ or $2m$, depending on how Limit is implemented. For example, if we pose the query "SELECT * FROM Employees Limit 10" and the system returns 10 arbitrarily chosen records, then if the employees table is modified by adding one record, the query could return 10 completely different records (for a symmtric difference of $20$). On the other hand, if the system returns records the $m$ in order based on their timestamp, then the symmetric difference still has size $2m$ (since the ordering information adds an implicit field). However, if the system returns the first $m$ records (based on timestamp) but randomly shuffles their order, then the size of the symmetric difference is $\min(2k, 2m)$ as adding/removing $k$ records can affect the result by at most $\min(2k, 2m)$. Due to these tricky complications, it is not recommended that a system provide an operator such as Limit.
    \item Bernoulli random sampling. If we take a table and perform a Bernoulli random sample (drop each record independently with probability $(1-p)$) then we require a more general notion of stability that accounts for the randomness \citep{ebadistability}. However, this is a stable operation, the effective size of the symmetric difference remains unchanged. Hence, this is a good replacement for Limit.  
    \item The Order By operator, which sorts the records in a table, is particularly troublesome since adding one record (which would be first in the sorted order) could change the positions of all other records. This introduces an implicit ordering field which causes the size of the symmetric difference to increase to the maximum table size. Most differentially private computations have no need for ordering (for example, if we need the average salary or the 95th percentile, we do not need to sort the data first) and hence it is not recommended that a system provide an operator such as Order By.
    \item SELECT/WHERE. Simple Select-Project queries of the form ``SELECT list-of-attributes FROM MyTable'' keep the size of the symmetric difference unchanged.
    \item Distinct. Transformations of the form SELECT DISTINCT(column-name) FROM MyTable'' also keep the size of the symmetric difference unchanged.
\end{itemize}

It is important to note that in special cases, some operations can actually decrease the size of the set difference, depending on the previous operations. For instance, going back to our motivating example at the beginning of Section \ref{subsec:stability}, if we added the transformation ``SELECT DISTINCT(Salary, Employee\_id) FROM Temp5'' then the entire sequence of transformations up to that point leaves the size of the symmetric difference unchanged (i.e. adding/removing $k$ people from the Employees table causes the output of the query sequence to change by at most $k$). Such special cases should be tested separately in the same was as for individual transformations (that is, the sequence of operations is run on two different inputs whose symmetric difference is known; the test measures the size of the symmetric difference of the outputs and compares it to the upper bound on the symmetric difference produced by the system).

In addition to tests, it is important to have a document that states and proves the correctness of the stability computations for each transformation.

\subsubsection{Tracking Sensitivity and Lipschitz Continuity}

For functions whose inputs are datasets and outputs are numerical vectors, the analogue of stability is \emph{sensitivity} which measures the following. If a total of $K$ records are added to/removed from the input then by how much (measured as $L_1$ distance) does the output change? One example is a function that computes the total salary of an input table. If $k$ records are added/removed, then the output changes by at most $k$ times the upper bound on salary. For some functions, the input is a vector and the output is a vector. For such functions, one can track the Lipschitz continuity -- if the input to a function changes by at most $c$ (in the $L_1$ distance), how much does the output change by? For example, if we have a function that takes as input a number and then doubles it, then the output changes by at most $2c$. Keeping track of sensitivity and Lipschitz continuity allows a system to determine the overall sensitivity of a calculation. For instance, suppose salary is capped by $300,000$. Then, the query "SELECT 2*SUM(Salary) FROM Employees" can be thought of as a sequence of two operations: compute the total salary then double it. Adding or removing one record changes the total by at most $300,000$ and the doubling operation turns that into $600,000$. Thus the sensitivity of the combined sequence of operations is $600,000$.

Sensitivity and Lipschitz continuity can be tested by feeding different inputs into a function and measuring how far apart the outputs are. Again, a document that states and proves the correctness of sensitivity and Lipschitz continuity calculations is necessary. These computations can be embedded into a type system that automatically tracks the overall sensitivity of a sequence of operations. For details, see the languages proposed by \cite{Fuzz,AFuzz,Dfuzz}.

\section{Summary and Conclusions}\label{sec:summary}
Differential privacy is a worst-case, mathematical statement about an algorithm, and as such can only be established through mathematical proof. But differential privacy is typically proved in an idealized setting based on mathematical abstractions of algorithms, rather than their concrete implementations, and this can introduce a significant challenge for deployed systems:
\begin{enumerate}
    \item Complicated algorithms have complicated analyses, and even the paper proofs of correctness in the idealized model may have errors.
    \item As with any large system, bugs can be introduced when translating the mathematical abstraction of the algorithm into code, and these bugs can invalidate the privacy guarantees.
    \item There are fundamental differences between the idealized model and implemented software: in practice, we do not have access to truly random numbers, continuous distributions, or infinite precision calculation. There are also side channels like run-time that are not typically considered in the idealized model.
\end{enumerate}

All of these issues make both careful \emph{code review} and \emph{testing} essential for deployed systems claiming differential privacy. Testing is not a substitute for mathematical proof, but rather a way of potentially detecting either errors in the mathematical proof or important gaps between the idealized model and the implemented system. These tests should be automated and periodically run so that errors introduced as new updates are deployed can be caught quickly.

Testing for differential privacy is a difficult task in and of itself, but can be simplified by a careful system design. The one we propose has two main features that aid in testing:
\begin{enumerate}
    \item It partitions functions across layers according to what needs to be tested to guarantee differential privacy: i.e. at the data access layer, it is deterministic sensitivity calculations. At the privacy layer, it is claimed properties of the distribution output by randomized functions. This modularity means, for example, that when testing functionality at the privacy layer, it is not necessary to run the code at the data access layer, which might be time consuming --- especially since tests at the privacy layer may need to run millions of times to guarantee sufficient statistical significance.
    \item It aims at keeping the ``privacy core'' as small as possible by pushing as much code as possible to the post-processing layer, whose correctness is immaterial for claims of differential privacy. This limits the quantity of code that needs to be tested in order to build confidence in a claimed degree of differential privacy.
\end{enumerate}
When possible, we also recommend that the core code on which privacy guarantees rely be made open source: this enables a wider population to verify correctness and discover and remediate bugs, and ultimately will help build confidence in the system.

Our guidelines make large-scale deployment of differentially private APIs feasible now. But there is need for more research on automatic verification and testing for differential privacy to make the process of development easier and less error prone. The automated "unit tests" that we propose in these guidelines are statistical hypothesis tests aimed at falsifying claims of worst-case differential privacy. However one can imagine automated tests that combine statistical and mathematical reasoning that aim to positively establish weaker notions, like ``Random Differential Privacy'' \cite{randomDP}. Weaker guarantees of this sort fall short of the end goal of differential privacy, but automated tests that can positively establish such guarantees complement statistical tests aimed at falsifying differential privacy to allow system builders to establish trust in their system.

\section{Acknowledgments}
We thank Mason Hemmel and Keegan Ryan from NCC group for pointing out additional sources of timing attacks on differential privacy systems.

\bibliographystyle{apalike}
\bibliography{references}

\end{document}